# Spin-Dependent Scattering off Neutral Antimony Donors in $^{28}$Si Field-Effect Transistors

C. C. Lo, J. Bokor

Department of Electrical Engineering and Computer Sciences, University of California,

Berkeley, CA 94720, USA

T. Schenkel

Accelerator and Fusion Research Division, Lawrence Berkeley National Laboratory,

Berkeley, CA 94720, USA

A. M. Tyryshkin, S.A. Lyon

Department of Electrical Engineering, Princeton University, Princeton, NJ 08544, USA

We report measurements of spin-dependent scattering of conduction electrons by neutral donors in accumulation-mode field-effect transistors formed in isotopically enriched silicon. Spin-dependent scattering was detected using electrically detected magnetic resonance where spectra show resonant changes in the source-drain voltage for conduction electrons and electrons bound to donors. We discuss the utilization of spin-dependent scattering for the readout of donor spin-states in silicon based quantum computers.

Silicon based quantum computation has attracted much interest since its original proposal by Kane [1]. Donor atoms (e. g. phosphorus [1] or antimony [2, 3]) embedded in a silicon substrate are the basis for quantum bits (qubits), and spins of donor electrons and nuclei are utilized for quantum information storage and manipulation. An integral part of any quantum computation architecture is a high-fidelity qubit readout. While many readout proposals have emerged in the past years [1, 4-7], experimental demonstration of spin-state detection of single donors has remained elusive. In this article we demonstrate a possible route towards single-spin detection for donor qubits based on spin-dependent scattering (SDS) of conduction electrons by neutral donors. At cryogenic temperatures the dominant scattering mechanisms of conduction electrons (or the two-dimensional electron gas, 2DEG) in metal-oxide-semiconductor devices include surface roughness scattering, charged defect scattering, and neutral impurity scattering [8]. Neutral impurity scattering is spin-dependent because different spin configurations of the conduction and donor electrons (singlet or triplet) imply a different spatial distribution of the two-electron wavefunction, which translates into a difference in scattering cross-sections. This SDS process by phosphorus impurities in an accumulation-mode field-effect transistor (aFET) was first observed by Ghosh and Silsbee using electrically detected magnetic resonance (EDMR) [9]. In an EDMR experiment, a static magnetic field induces a Zeeman splitting in the electron energy levels, and in thermal equilibrium, triplet scattering is favored as more spins are aligned with the static field. Singlet scattering can be enhanced by inducing spin flips with a resonant microwave field. This increase in singlet content then registers as an effective channel resistance change of the aFET. Ghosh and Silsbee used large-area aFETs ($1\times0.1$ mm$^2$) formed in bulk-doped silicon with about $2\times10^{17}$ phosphorus/cm$^3$ [9]. The number of donors close (~10 nm) to the aFET channel that contribute to SDS was estimated to be ~$10^8$. However, bulk donors far away from the channel that did not contribute to SDS caused an undesired bolometric signal due to resonant microwave absorption, and substantial efforts were undertaken to resolve interfering bolometric effects and to isolate the SDS signal.

In the present work, we demonstrate SDS by neutral $^{121}$Sb donors in silicon aFETs. In order to avoid bolometric signals, aFETs were formed in undoped silicon and ~$6\times10^6$ donors were implanted into the transistor channel. While most donor-based silicon quantum computer proposals have suggested spins of $^{31}$P as qubits, $^{121}$Sb is used in our experiments due to its smaller straggling in the channel implantation process, lower diffusion rates in silicon, and to avoid spurious signals arising from residual background $^{31}$P atoms in the silicon substrate or from the polycrystalline silicon gate. Moreover, electron spin relaxation rates of implanted $^{121}$Sb and the Stark effect due to applied electric fields have been previously studied in detail [2, 3]. In the limit of a single-donor doped aFET, an EDMR experiment can yield spectra where information on a single nuclear spin state can be deduced from the presence (and absence) of donor hyperfine-split peaks, provided that the read-out time is faster than the spin-flip time of the nuclear spin.

AFETs were fabricated in isotopically enriched $^{28}$Si epi-layers (2 μm thick, >99.9% enrichment) on undoped, natural silicon (100) substrates. The channel area ($160\times20$ μm$^2$) was implanted with $^{121}$Sb at 80 keV and a dose of $2\times10^{11}$ /cm$^2$. Subsequently, a 20 nm gate oxide was grown and in-situ phosphorus-doped polycrystalline silicon was deposited and patterned as the gate electrode. Arsenic was then implanted ($5\times10^{15}$ /cm$^2$, 40 keV) to form degenerately doped source-drain regions (Figure 1(a)). A forming gas anneal at 400°C for 20 minutes was performed to passivate defects at the Si/SiO$_2$ interface. The post-processing peak dopant concentration is about $3\times10^{16}$ /cm$^3$, ~30 nm below the oxide interface, as determined from Taurus TSUPREM-4 simulations and Secondary Ion Mass Spectrometry measurements. The threshold voltage, $V_t$, of the aFETs was 0.25 V at 5 K.

EDMR was performed with a modified X-band (9.6 GHz) ESR spectrometer (Bruker Elexsys 580). A continuous microwave excitation at constant frequency and power was applied, and the DC magnetic field ($B_0$) was scanned with an aFET accurately positioned inside a cylindrical microwave resonator. The source-drain channel of the aFET is oriented along the symmetry axis of the resonator, parallel to the magnetic component ($B_1$) and perpendicular to the electric component ($E_1$) of the microwave field (Figures 1(b), (c)). In order to minimize microwave absorption by metallic parts of the device, we adopted an elongated chip layout [10]. The device was current biased through the source and drain terminals, and the drain-source voltage ($V_{ds}$) was monitored while the $B_0$ field was swept. The gate voltages ($V_g$) and channel currents ($I_{ds}$) were chosen such that aFETs operated in the linear regime to ensure a uniform 2DEG density throughout the channel. We used magnetic field modulation at 1 kHz with peak-to-peak amplitude $B_{mod} = 0.2$ mT to enhance the signal-to-noise ratio. All measurements were performed at 5 K.

The resonance condition for 2DEG electrons is given by the Zeeman splitting, $h\nu = g_{ce}\mu_B B_0$, where $g_{ce}$ is the conduction electron g-factor for silicon, $\mu_B$ the Bohr magneton, and $h$ the Plank constant. For donor electrons, the resonance condition in first approximation is $h\nu = g_{de}\mu_B B_0 + Am_I$, where $g_{de}$ is the donor electron g-factor which is slightly shifted from $g_{ce}$ due to enhanced spin-orbit interaction. The hyperfine interaction with donor nuclei, $A$, introduces additional splittings, and six transitions are expected for the nuclear spin projections, $m_I$, of $^{121}$Sb (nuclear spin $I = 5/2$). From Ref. 9, the EDMR signal amplitude for SDS of 2DEG electrons off donors can be described as:

$$\frac{\Delta R}{R} = \alpha \left[ P_{ce}^0 P_{de}^0 (1 - (1 - s_{ce})(1 - s_{de})) \right] \kappa, \qquad (1)$$

where $R$ is the aFET channel resistance in thermal equilibrium, and $\Delta R$ the change in channel resistance upon excitation by spin resonance transitions. $\alpha = \langle \Sigma_S - \Sigma_T \rangle / \langle \Sigma_S + 3\Sigma_T \rangle$ is the normalized difference of singlet ($\Sigma_S$) and triplet ($\Sigma_T$) scattering cross-sections. $P_{ce}^0$ and $P_{de}^0$ are the thermal equilibrium polarizations of the 2DEG and donor electrons, respectively. $s_{ce}$ and $s_{de}$ are the microwave saturation parameters for the 2DEG and donor electrons. Equation (1) is symmetric with respect to 2DEG and donor electrons when the applied microwave power is sufficiently large to saturate both spin transitions. Thus when $s_{ce} = s_{de} = 1$ EDMR signals for 2DEG and donor electrons are expected to have a ratio of $(2I+1):1$. $\kappa$ (<1) is a device-dependent parameter that weights the contribution from SDS against other scattering processes in the device.

Since a field-modulation technique was used and the field modulation amplitude was smaller than the spectral line widths, the measured EDMR signal can be approximated as a first derivative signal $[d(\Delta V_{ds}/V_{ds})/dB_0]B_{mod}$. Figure 2(a) shows the EDMR spectrum of an aFET, where we have translated the raw EDMR data to $[d(\Delta R/R)/dB_0]B_{mod}$. The strong central peak is from the 2DEG as evidenced by its g-factor $g_{ce}$= 1.9998 [11, 12]. Six weaker peaks are from $^{121}$Sb donors and correspond to the six nuclear spin projections. Using the Breit-Rabi formula, the g-factor $g_{de}$= 1.9985(1) and the hyperfine coupling constant $A = 6.62$ mT can be extracted from the peak positions and are found to be in good agreement with published data for $^{121}$Sb in silicon [13]. The spectrum in figure 2(a) was taken with an applied power of 2.53 mW, which is in the weakly saturated regime (see below). The ratio of signal amplitudes for conduction electrons vs. donors is about 20:1, while a ratio of 6:1 is expected if neutral impurity scattering off $^{121}$Sb atoms was the only SDS process involved. We speculate that spin-dependent electron-electron scattering, similar to the case in Si/SiGe heterostructures [12], is responsible for the enhanced 2DEG signal.

Both the 2DEG and donor signals decrease with increasing gate voltage (Figure 2(b), (c) and 3(a)). This is because the spin polarization of conduction electrons decreases as $1/(V_g - V_t)$ [9]. Moreover, as $V_g$ is increased, the conduction electron wavefunction is more confined at the Si/SiO$_2$ interface, hence fewer donors contribute to the SDS signal as donors further from the interface no longer interact with the 2DEG. This gate voltage dependence strongly suggests that the EDMR signal is due to SDS rather than bolometric effects involving donors far from the channel. Measurements with different drain currents (0.8 to 1.6 µA) showed no effect on the signal amplitude, which implies that Joule heating in the channel is negligible at these current densities.

The microwave power dependence of the EDMR signal is shown in Figure 3(b). Both signals saturate at high microwave power, as expected from SDS processes. The donor signal saturates at slightly lower power than the 2DEG signal, indicating longer relaxation times ($T_1$) for donor spins. The exact magnitude of microwave fields in our sample is not known since electrical leads can act as antennas and enhance local fields significantly [14]. The below-saturation peak-to-peak line widths are found to be $0.26 \pm 0.02$ mT for 2DEG electrons and $0.20 \pm 0.02$ mT for donor electrons, at $V_g = 0.45$ V. We note that these line widths are larger than expected for donors in a nuclear-spin free environment [2]. The signal line shapes are not simple Lorentzians at low modulation amplitudes and we speculate that inhomogeneous broadening played a significant role.

Two critical characteristics for qualification of SDT as a mechanism for readout of single nuclear spin states are signal amplitudes, $\Delta R/R$, and spin relaxation times during the readout process. The maximum signal amplitudes of resonant current changes for donors we observed were $\sim 10^{-7}$ of the off-resonant current (at $V_g = 0.35$ V). It is, however, not known how many of the $6 \times 10^6$ donors in the channel actually contribute to our EDMR signal. Doubly-occupied $D^-$ states can form for donors close to the interface, and it is not known up to which depth neutral donors can have sufficient overlap with the wavefunctions of conduction electrons in the shallow 2DEG to contribute to the signal. Electron spin relaxation times of implanted donors, $T_{1e}$, at 5 K are ~15 ms [2], and nuclear spin relaxation times, $T_{1n}$ are at least 300 times longer [15]. But $T_{1e}$ and $T_{1n}$ during EDMR measurements are not yet known. Typical line widths in our experiments were ~0.2 mT, yielding an estimate of a lower bound for $T_{1e}$ during readout of >0.1 μs. With a readout current of ~1.6 μA, this allows collection of >$10^6$ ( >$3\times10^8$ ) electron charges within $T_{1e}$ ($T_{1n}$). SDS is a potential mechanism for readout of single nuclear spin states, but spin relaxation times have to be quantified, and devices have to be optimized for enhanced signal amplitudes.

In conclusion, we have observed spin-dependent scattering of conduction electrons off neutral donors by electrically detected magnetic resonance in aFETs formed in $^{28}$Si. Resonance signals of 2DEG electrons and hyperfine-split peaks from bound electrons of channel-implanted $^{121}$Sb donors are detected. The high sensitivity of EDMR enables studies of small spin ensembles [16] and promises to allow scaling to the few and single-donor regime with optimized devices for readout of single nuclear spin states in qubit donors.

**ACKNOWLEDGEMENTS**

The authors thank R. de Sousa for helpful discussions. This work was supported by the National Security Agency under MOD 713106A, the Department of Energy under Contract No. DE-AC02-05CH11231, the National Science Foundation under Grant No. 0404208, and the Nanoelectronics Research Initiative-Western Institute of Nanoelectronics. Support in device fabrication by the UC Berkeley Microlab staff is gratefully acknowledged.

**REFERENCES**

1	B. E. Kane, Nature 393, 133-137 (1998)

2	T. Schenkel, J. A. Liddle, A. Persaud, A. M. Tyryshkin, S. A. Lyon, R. de Sousa, K. B. Whaley, J. Bokor, J. Shangkuan, and I. Chakarov, Appl. Phys. Lett. 88, 112101 (2006)

3	F. R. Bradbury, A. M. Tyryshkin, G. Sabouret, J. Bokor, T. Schenkel, and S. A. Lyon, Phys. Rev. Lett. 97, 176404 (2006)

4	M. J. Testolin, A. D. Greentree, C. J. Wellard, and L. C. L. Hollenberg, Phys. Rev. B 72, 195325 (2005); A. D. Greentree, A. R. Hamilton, L. C. L. Hollenberg, and R. G. Clark, Phys. Rev. B 71, 113310 (2005)

5	A. G. Petukhov, V. V. Osipov, and V. N. Smelyanskiy, Appl. Phys. Lett. 89, 153127 (2006)

6	A. Yang, M. Steger, D. Karaiskaj, M. L. W. Thewalt, M. Cardona, K. M. Ithoh, H. Riemann, N. V. Abrosimov, M. F. Churbaniv, A. V. Gusev, A. D. Bulanov, A. K. Kaliteevskii, O. N. Godisov, P. Becker, H. J. Pohl, J. W. Ager, E. E. Haller, Phys. Rev. Lett. 97, 227401 (2006)

7	A. R. Stegner, C. Boehme, H. Huebl, M. Stutzmann, K. Lips, and M. S. Brandt, Nature Physics 2, 835 (2006)

8	T. Ando, A. B. Fowler, and F. Stern, Rev. Mod. Phys. 54, 437 (1982)

9   R. N. Ghosh and R. H. Silsbee, Phys. Rev. B 46, 12508 (1992)

10  C. Boehme and K. Lips, Physica B 376-377, 930 (2006)

11  C. F. Young, E. H. Poindexter, G. J. Gerardi, W. L. Warren, and D. J. Keeble, Phys. Rev. B 55, 16245 (1997)

12  C. F. O. Graeff, M. S. Brandt, M. Stutzmann, M. Holzmann, G. Abstreiter, and F. Schaffler, Phys. Rev. B 59, 13242 (1999)

13  G. Feher, Phys. Rev. 114, 1219 (1959); J. Eisinger and G. Feher, Physical Review 109, 1172 (1958)

14  G. Kawachi, C. F. O. Graeff, M. S. Brandt, and M. Stutzmann, Jap. J. Appl. Phys. 36, 121 (1997)

15  A. M. Tyryshkin, J. J. L. Morton, A. Ardavan, S. A. Lyon, J. Chem. Phys. 124, 234508 (2006)

16  D. R. McCamey, H. Huebl, M. S. Brandt, W. D. Hutchison, J. C. McCallum, R. G. Clark, and A. R. Hamilton, Appl. Phys. Lett. 89, 182115 (2006); M. Xiao, I. Martin, E. Yablonovitch, and H. W. Jiang, Nature 430, 435 (2004)

**Figure captions**

Figure 1. (a) Schematic cross-section of an aFET. (b) Device placement and field orientations in the ESR microwave resonator. (c) Magnified view of an aFET chip.

Figure 2. (a) EDMR spectrum from an $^{121}$Sb-doped aFET at 5K ($I_{ds}$ = 1.58 µA, $V_g$ = 0.45V). (b) EDMR spectra for a series of gate voltages. Only the inner two donor hyperfine-split peaks are shown for clarity. (c) The same EDMR spectra as (b) with the y-axis magnified tenfold to highlight the $^{121}$Sb peaks.

Figure 3. (a) Gate voltage dependence of EDMR signals ($P_{mw}$ = 3 mW, $I_{ds}$ = 0.8 µA). (b) Microwave power dependence of EDMR signals ($V_g$ = 0.45 V, $I_{ds}$ = 0.8 µA).

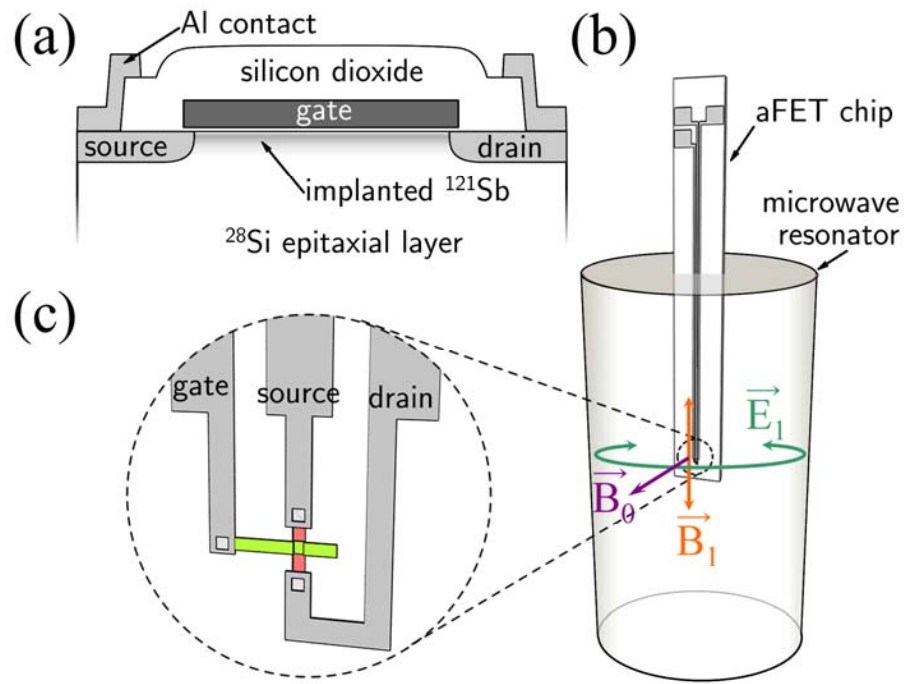

Figure 1

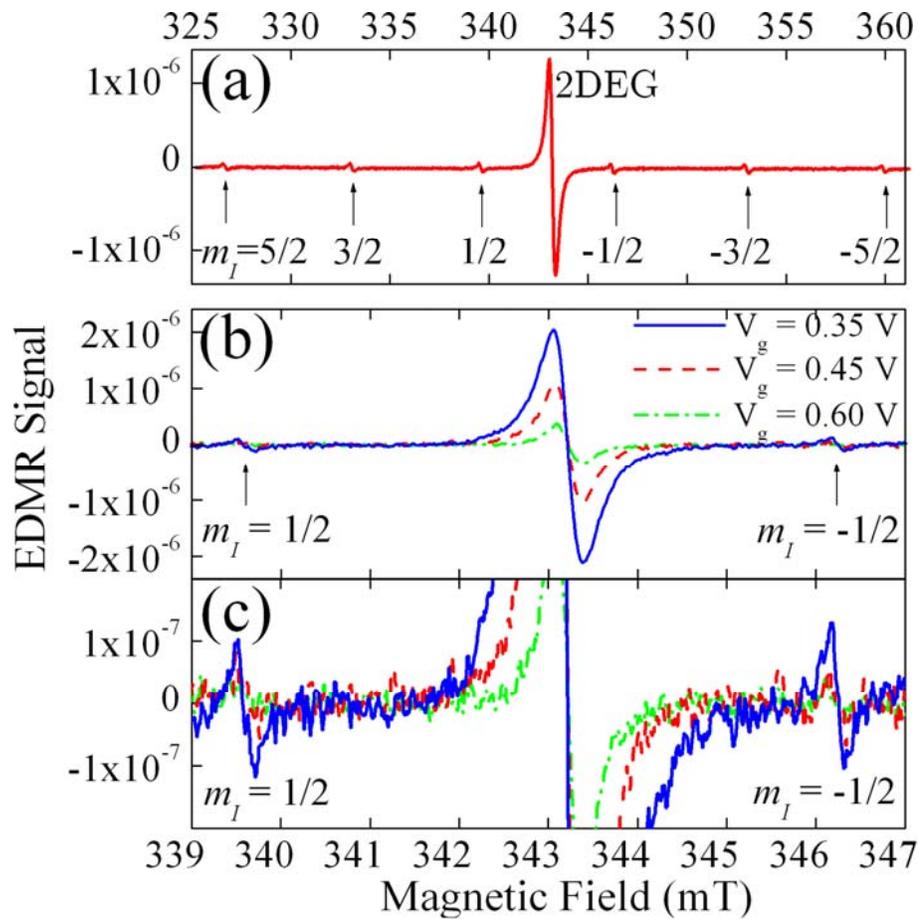

Figure 2

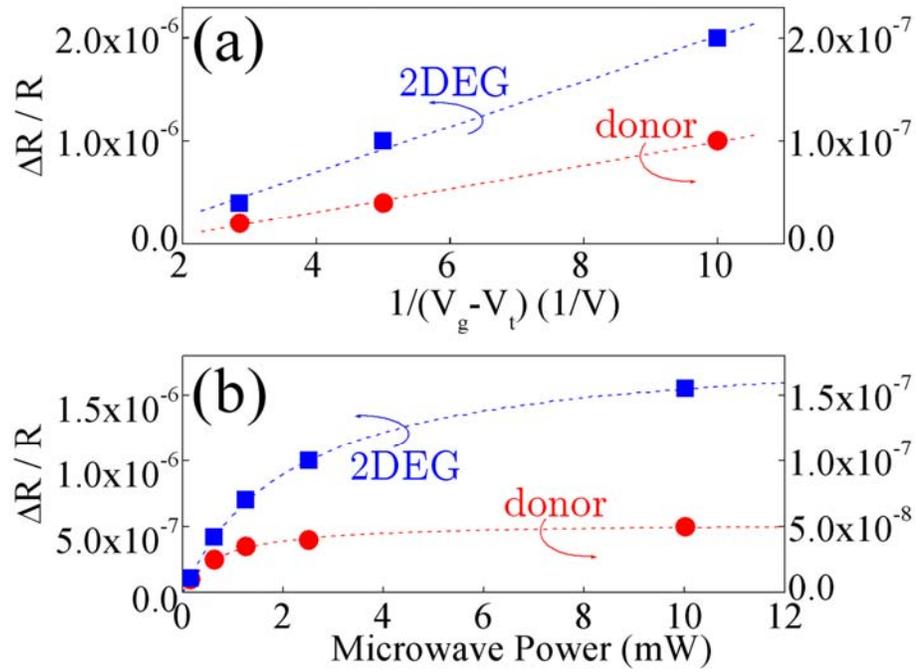

Figure 3